\documentclass[a4paper,11pt]{article}
\usepackage{pos}
\usepackage{setspace}

\title{First follow-up of transient events with the CTA Large Size Telescope prototype}
 \ShortTitle{First transient observations with LST-1}

\author*[a]{Alessandro Carosi}
\author[b]{Halim Ashkar}
\author[c]{Alessio Berti}
\author[d]{Pol Bordas}
\author[e]{Mathieu de Bony Lavergne}
\author[f]{Alice Donini}
\author[a]{Mykhailo Dalchenko}
\author[e]{Armand Fiasson}
\author[a]{Luca Foffano}
\author[g]{Satoshi Fukami}
\author[g]{Yukiho Kobayshi}
\author[b]{Francesco Longo}
\author[g]{Koji Noda}
\author[e]{David Sanchez}
\author[e]{Monica Seglar-Arroyo}
\author[b]{Fabian Sch\"ussler}
\author[e]{Thomas Vuillaume}
\affiliation[a]{DPNC – University of Geneva 24 quai Ernest Ansermet, CH1211 Genève 4, Switzerland}
\affiliation[b]{IRFU/DPhP, CEA, Universit\'e Paris-Saclay, F-91191 Gif-sur-Yvette, France}
\affiliation[c]{Max-Planck-Institut für Physik, Föhringer Ring 6, 80805 München, Germany}
\affiliation[d]{IEEC-UB, 08028, Barcelona, Spain}
\affiliation[e]{LAPP, Universit\'e de Savoie Mont-Blanc, CNRS/IN2P3, 74941, Annecy, France}
\affiliation[f]{Institut de Fisica d’Altes Energies, The Barcelona Institute of Science and Technology, Bellaterra, Spain}
\affiliation[g]{Institute for Cosmic Ray Research, The University of Tokyo, Kashiwa, Japan}
\affiliation[h]{INFN Trieste and Università degli Studi di Trieste, 34127 Trieste, Italy}

\forColl{CTA LST} 
\emailAdd{alessandro.carosi@unige.ch}
\abstract{
The recent detection of a very high energy (VHE) emission from Gamma-Ray Bursts (GRBs) above 100 GeV performed by the MAGIC and H.E.S.S. collaborations, has represented a significant, long-awaited result for the VHE astrophysics community. Although these results' scientific impact has not yet been fully exploited, the possibility to detect VHE gamma-ray signals from GRBs has always been considered crucial for clarifying the poorly known physics of these objects. Furthermore, the discovery of high-energy neutrinos and gravitational waves associated with astrophysical sources have definitively opened the era of multi-messenger astrophysics, providing unique insights into the physics of extreme cosmic accelerators. In the near future, the Cherenkov Telescope Array (CTA) will play a major role in these observations. Within this framework, the Large Size Telescopes (LSTs) will be the instruments best suited to significantly impact on short time-scale transients follow-up thanks to their fast slewing and large effective area. The observations of the early emission phase of a wide range of transient events with good sensitivity below 100 GeV will allow us to open new opportunities for time-domain astrophysics in an energy range not affected by selective absorption processes typical of other wavelengths. In this contribution, we will report about the observational program and first transients follow-up observations performed by the LST-1 telescope currently in its commissioning phase on La Palma, Canary Islands, the CTA northern hemisphere site.
}

\FullConference{37$^{\rm{th}}$ International Cosmic Ray Conference (ICRC 2021)\\
		July 12th -- 23rd, 2021\\
		Online -- Berlin, Germany}

\begin{document}
\maketitle

\section{Introduction}
\label{sec:intro}
Since more than 50 years from their discovery, gamma-ray bursts (GRBs) are still the targets of large observational programs in different energy bands by both ground-based and space-based instruments. In 2019, the first firm detection of a VHE gamma-ray emission component from GRBs has definitively opened a new observational window for the study of those enigmatic transient events. At present, a bunch of new detections have been announced with first results already published for a sub-sample of events like GRB~180720B~\cite{HESS_GRB180720B}, GRB~190114C~\cite{MAGIC_GRB190114C_1} and GRB~190829A~\cite{HESS_GRB190829A}. These discoveries represent the results of a $\sim$20-years-long-lasting hunt by the major Cherenkov telescope collaborations and they represent a remarkable step forward in our understanding of GRB physics. For long time, detecting a VHE signal associated with GRBs posed a major challenge for Imaging Atmospheric Cherenkov Telescope (IACTs) from both the technical and the scientific point of view (see e.g.~\cite{Covino10}~\cite{Carosi13}). On the other hand, the possibility to detect VHE gamma-ray signal from GRBs is crucial for clarifying the poorly-known physics of these objects during the different phases of their emission. This is particularly important during the early afterglow phase when the co-existence of forward and reverse shocks in the emitted outflow could result in a large variety of different emitting scenarios with overlapping emission components. GRBs have always been considered the prototype of cosmic transient objects. However, the recent growth of multi-messenger astrophysics opened the possibility to extend dedicated follow-up campaigns to alerts coming from different cosmic signal such as gravitational waves (GWs) and neutrinos (see e.g., \cite{Longo21,hess_gw1}). The discovery of the connection between GW transient signals and short-GRB~\citep{GW-sGRB} has indeed proven that GW astrophysical sources are related to extreme objects and environments that are also expected to emit photons and, possibly, neutrinos. Therefore, observations in the GW/neutrino and electromagnetic (EM) channels represent the way to reach a more complete comprehension of such astrophysical sources, their emission engines and the physics of their progenitors and their environment. The CTA is currently setting up dedicated follow-up programs of GRB/GW/neutrino alerts as well as other type of transient objects in the very high energy band~\cite{CTAmmessenger}.

Within the CTA framework, the LSTs are particularly suited for GRB and transient studies thanks to the fast repositioning speed and the low-energy threshold that reduces the effect of the flux attenuation by pair production with the lower energy (optical/IR) photons of the diffuse extragalactic background light (EBL).

\section{GRBs at VHE energy}
\label{sec:GRBatVHE}
According to the widely accepted relativistic shock model known as {\it fireball} (see e.g., \cite{Paczynski1986} \cite{Piran1999}), GRB emission arises from the conversion of the kinetic energy of a relativistic outflow into electromagnetic emission. Although the details of this conversion remain poorly understood, a largely discussed possibility is that the observed photons are generated by particles accelerated to ultra-relativistic energies by successive collisions within a magnetized medium. These particles can emit the observed high-energy photons by many possible non-thermal mechanisms. In particular, synchrotron emission has largely been considered as the most natural to explain the GRB sub-MeV emission \cite{SaEs01}\cite{ZM01}\cite{GZ07}. Although it cannot fully explain the observed prompt spectrum for the majority of the events, synchrotron emission is believed to play an essential role in GRB dynamics. In particular, it has been suggested that the GeV emission observed by {\it Fermi}-LAT extending after the end of the prompt emission is synchrotron radiation produced at the external shock that is driven by the jet into the circum-burst medium (see e.g.,~\cite{Ghisellini2010})
\begin{figure}
\centering
\includegraphics[width=0.485\columnwidth]{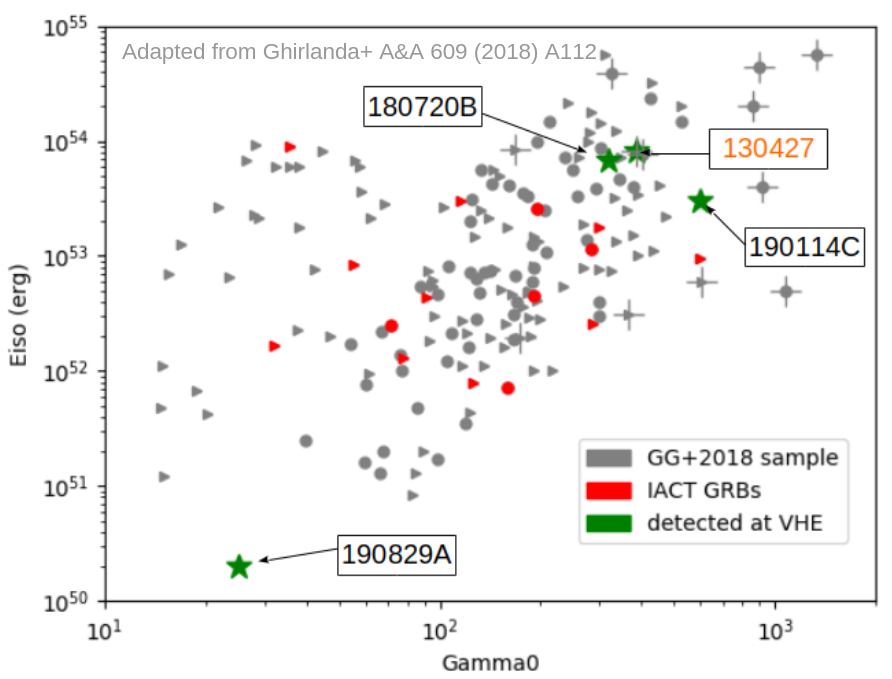}
\includegraphics[width=0.5\columnwidth]{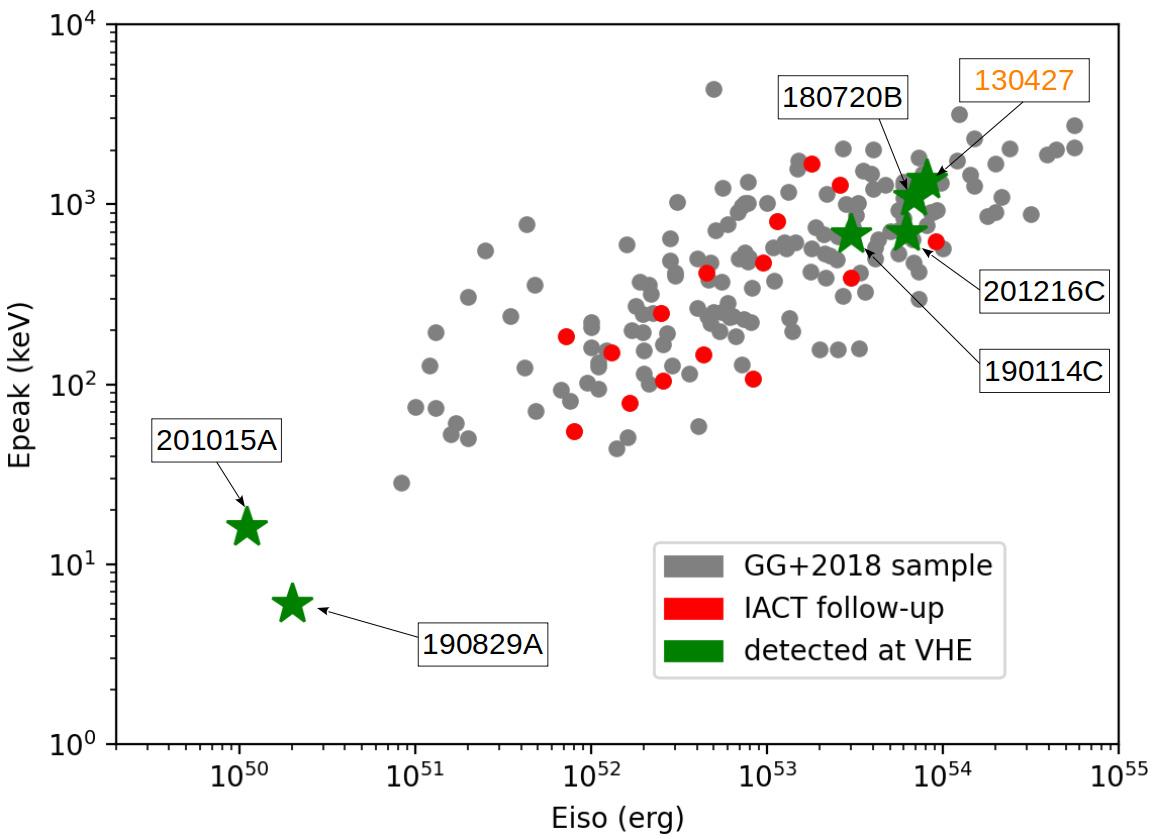}
\caption{{\it left panel:} Correlation between the bulk Lorentz factor at the beginning of the afterglow phase ($\Gamma_0$) and the isotropic equivalent energy $E_{iso}$ for the sample of GRB reported in~\cite{Ghirlanda2018}. {\it right panel:} the empirical correlation (Amati relation) between the isotropic equivalent energy $E_{iso}$ and the peak energy of the GRB spectrum for the same sample in~\cite{Ghirlanda2018} and for the events detected in the VHE band.}
\label{fig:amati}
\end{figure}
It is important to note that the detection up to TeV energies by MAGIC (GRB\,190114C) and H.E.S.S. (particularly GRB\,190829A), provided an unexpected complexity scenario due to the remarkable differences between the phenomenology of the two GRBs. In both cases, gamma rays of such high energies well exceed the maximum energy achievable with synchrotron implying the co-existence of an extra emission component in the VHE band~\cite{MAGIC_GRB190114C_1}. On the other hand, the events stand on the opposite edges of the GRB energy distribution being on the $\sim 30\%$ sub-sample of most energetic burst for GRB~190114C ($E_{iso} = 3 \times 10^{53}$ erg) and more than 3 orders of magnitude lower for GRB~190829A ($E_{iso} = 2 \times 10^{50}$ erg). Consequently, also some physical characteristic and observable result in a completely different distribution for the two events. As an example, Fig.~\ref{fig:amati} (left panel) shows the value of the bulk Lorentz factor ($\Gamma_0$) evaluated for a large sample of GRBs~\cite{Ghirlanda2018} once known their isotropic equivalent energy. The position of the confirmed VHE detected GRBs (and additionally the $\sim 100$ GeV event GRB~130427A detected by {\it Fermi}-LAT) is reported by the green stars. Although the number of detections still cannot allow a full, statistically-significant population study, the events detected so far appear to cluster in different regions of the considered phase space parameters. On the other hand, all the events lie on the Amati relation (Fig.~\ref{fig:amati} right panel~\cite{Martone2017}), a well-known GRB energy and luminosity empirical correlation. This likely implies that the observed differences in luminosity and energetics might not be related to diferences in the geometry of the emission for the considered GRBs. The two events also significantly differ in their temporal profile with an extreme bright VHE emission lasting $\sim 15$ minutes for GRB~190114C and a dimmer but much longer-lasting emission for GRB~190829A (up to few days after GRB onset). Whether the differences between the two GRBs are intrinsic or related to their different distance ($z=0.08$ and $0.425$ for GRB~190829A and GRB~190114C respectively) is still unclear. 

Thus, this emerging puzzling scenario shows the importance of keep observing GRBs in the VHE band with next-generation IACTs to investigate the parameter space of VHE-transient emitters and their characteristics.
%Thus, although important, those results show a still unsatisfactory level of comprehension of GRB physics and the importance of keep observing GRBs in the VHE band with next-generation IACTs to investigate the parameter space of VHE-transient emitters and their characteristics.

\section{The LST-1 prototype}
\label{sec:LST1}

\begin{figure}
\centering
\includegraphics[width=0.45\columnwidth]{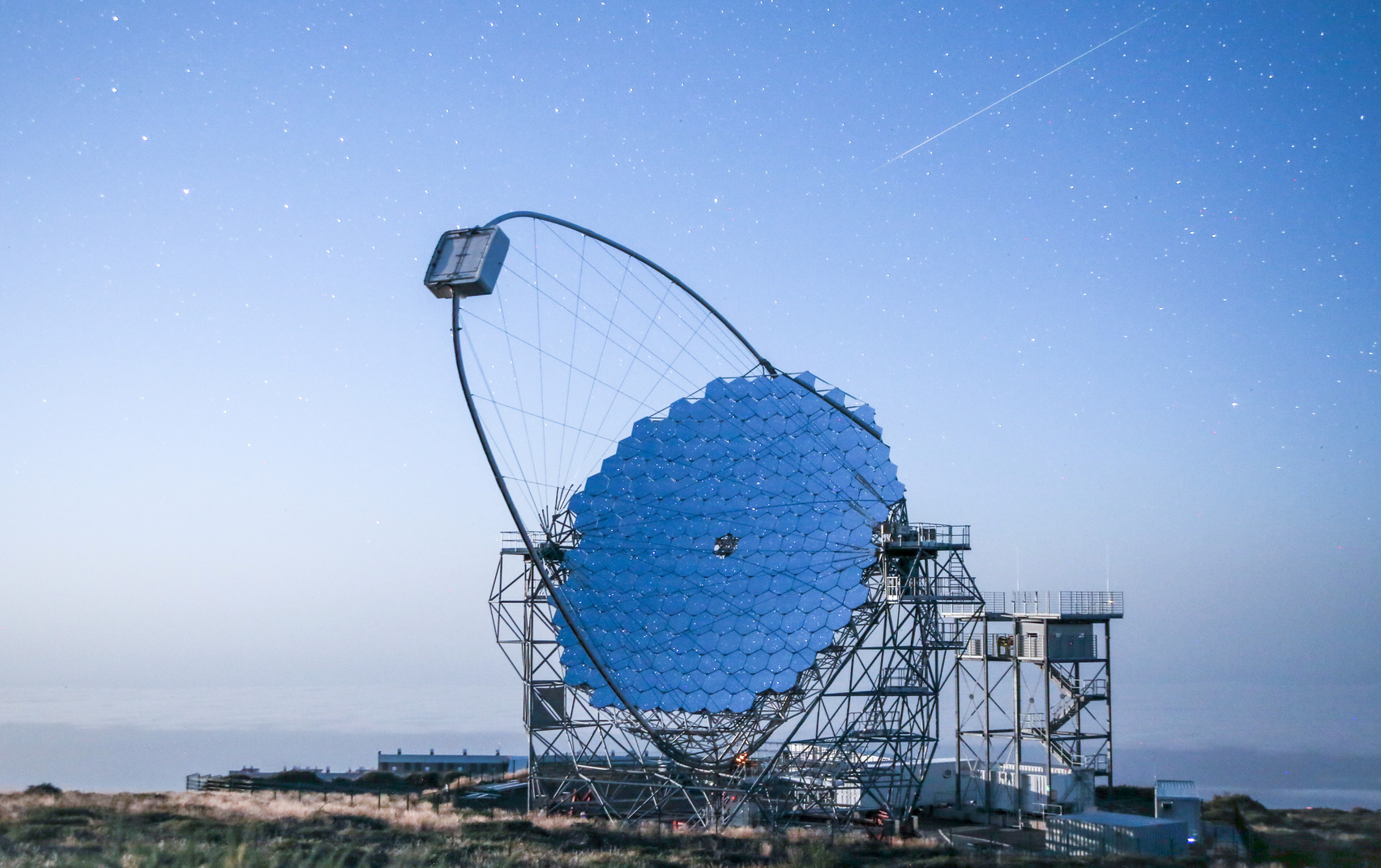}
\includegraphics[width=0.5\columnwidth]{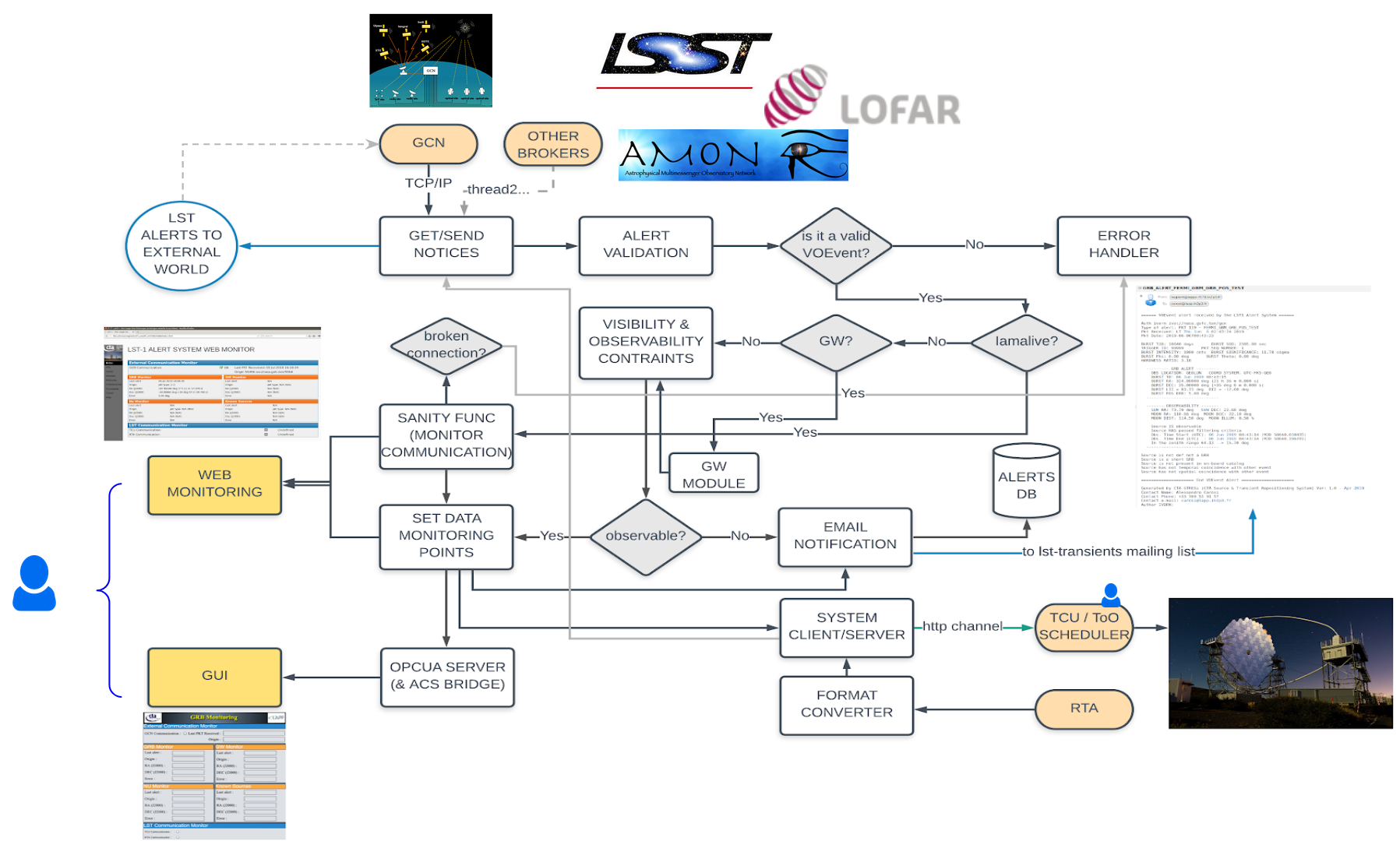}
\caption{The LST prototype during night operation in La Palma. Picture credit: Tomohiro Inada. Basic flowchart of LST-1 transient handler. }
\label{fig:lst1}
\end{figure}

The CTA represents the next generation ground-based observatory for the study of VHE gamma rays. It will consist of two arrays, instrumented with IACTs of different size and characteristic, one for each hemisphere. LSTs are the largest telescopes designed for CTA, having a 23~m diameter reflector. The first prototype, LST-1 (Fig.~\ref{fig:lst1} left panel), is located at the Roque de los Muchachos observatory ($28.8^{\circ}$~N, $17.8^{\circ}$~W, 2200 m a.s.l.), on the Canary Island of La Palma~\citep{lst-telescope-report-2019}. The telescope is equipped with a $\sim 4.5^{\circ}$ field-of-view camera composed by 1855 photo-multipliers (PMTs) tubes converting the Cherenkov light into electrical signals recorded by a fast readout system. Thanks to the wide reflective surface of about 400~m$^2$, the LST-1 will be able to achieve an energy threshold of 20~GeV, a value particularly suitable for transients and high-redshift sources observations. Furthermore, LSTs are built with a light carbon-fiber structure in order to reduce the total weight of the telescope to about 103 tons and to make possible the fast re-positioning ($\sim 30$~s for $180^{\circ}$ azimuth displacement) to catch early emission phases of transient objects. LST-1 was inaugurated in October 2018 and is currently finalizing its commissioning phase. 
%During this commissioning period, the telescope is  tested and provided with the most updated technologies concerning software and hardware in order to accomplish the strict requirements expected by CTA Observatory.
\subsection{The LST-1 transient handler}
The LST-1 response to external triggers relies on a specific {\it transient handler} system receiving the external trigger provided by the GRB Coordinate Network (GCN\footnote{http://gcn.gsfc.nasa.gov/}) through TCP/IP socket and with the following baseline functionalities:

\begin{itemize}
\item handle the communication with external resources according to their specific communication protocols (VOEvents~\cite{VOE}, binary socket, e-mail...);
\item handle the incoming alerts: receive, parse and archive relevant information;
\item check for visibility and/or filtering of the incoming alert according to pre-defined observational strategy and scheduling;
\item handle internal communication with the relevant telescope's sub-system as Telescope Control Unit (TCU), scheduler and Real Time Analysis (RTA);
\end{itemize}
The basic flowchart of the LST-1 transient handler is reported in the left panel of Fig.~\ref{fig:lst1}. The non-stopping and efficient connection with external facilities will provide the possibility to perform rapid follow-up on wide range of astrophysical sources like GRBs, galactic transients and the possible VHE electromagnetic counterparts of neutrinos and gravitational waves. To this end, the implementation of an efficient observational strategy based on specific science cases is also needed to define both the selection criteria for the different type of incoming alerts and the timing for the follow up observation within the general observation schedule of the telescope. Furthermore, it is currently under study the possibility to interface the LST-1 alert system with its RTA guarantying the possibility not only to receive but also to {\it deliver} alerts on possible VHE transients to the external astrophysical community almost in real-time. This will represent a major step forward for IACTs and a noticeably test-bed for the full-configured CTA.

\subsubsection{Large-error localization alerts}
\begin{figure}
\centering
\includegraphics[width=0.49\columnwidth]{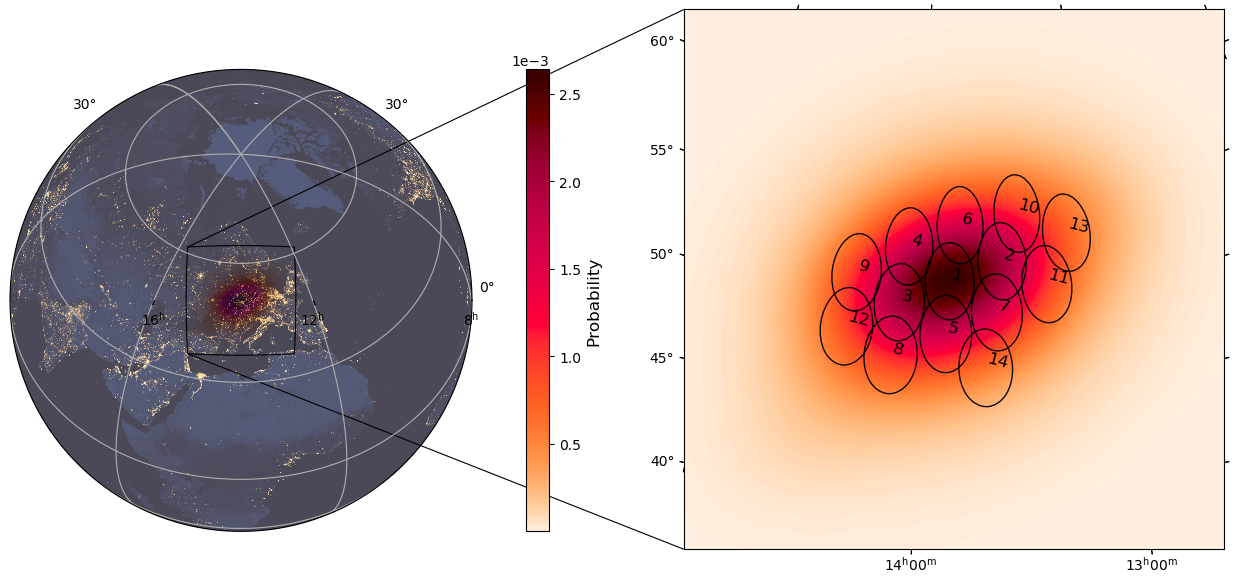}
\includegraphics[width=0.49\columnwidth]{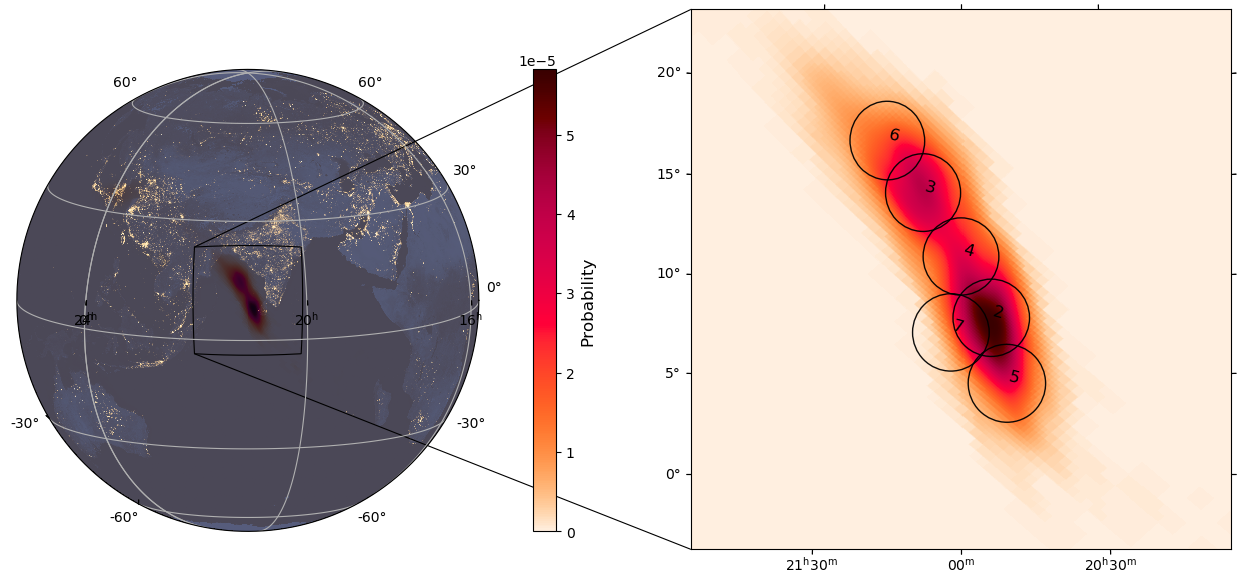}
\caption{Simulated follow-up of the GRB\,200303A GBM alert (left panel)~\citep{GCN27294} and the GW event, GW190915\_235702~\citep{O2_paper_catalog} (right panel). The skymaps represent the coverage of the GW and GRB localisation region as derived by the LST transient handler and considering a telescope's Field of View (FoV) of $2^{\circ}$. The scheduled observations are showed in chronological order. The achieved coverage is 63\% and 52\% for the two cases respectively. The Earth is shown in the background at the time of the start of observations.}
\label{fig:GW190915_235702_OBS_PrettyMap}
\end{figure}

Optimized pointing strategies have been mainly developed by mid- and small- FoV instruments in the context of gravitational wave follow-up campaigns, and actively used in current generation IACTs \citep{HESSrapid}\citep{BBHHESS}. The localization uncertainty region of the source emitting the detected gravitational waves can reach enormous sizes, ranging from 10-1000 deg$^2$, which is a major challenge to follow-up observatories which aim to observe the electromagnetic counterpart.  Built on the success of these strategies to cover large regions of the localization uncertainty region, these have been included in the LST-1 transient handler and broaden to various types of events with large localization uncertainties. These optimized observation strategies have been adapted to follow  gravitational wave alerts during the upcoming LIGO-Virgo-KAGRA 1-year-observation run O4, for which an improvement to 33$^{+5}_{-5}$ deg$^2$ for the mean 90\% location uncertainty for BNS is expected while a total of 10$^{+52}_{-10}$ BNS detections for the entire run is foreseen \cite{ProspectsGW}. However, the same algorithms might be used for other type of not-well-localized triggers such as GRB alerts from {\it Fermi}-GBM as well as neutrino alerts from IceCube, Antares and KM3Net in the future. An example of the observation strategy for a GW and a GBM alert is presented in Figure~\ref{fig:GW190915_235702_OBS_PrettyMap}. 

\section{First transients follow-up with LST-1}
\label{sec:firstfollowup}
At the time of writing, LST-1 prototype is finalizing its commissioning phase. However, starting from the first months of 2021, the time allocated for technical observations has been gradually reduced allowing the first observations of targets of astrophysical interest. Transients and, in particular GRBs, follow-up have the highest priority among LST-1 observed targets. Although a fully automatic procedure that will allow the telescope to react automatically to incoming alerts is still under development, a manual human-in-the-loop reaction can take place to start follow-up observation. Unfortunately, the beginning of the regular follow-up operations also coincided with a mentioned malfunctioning of the {\it Swift} satellite. The list of transient events observed between December 2020 and for first 6 months of 2021, is reported in Tab.~\ref{tab:transients}           

\begin{table}[!h]
\centering
\begin{center}
\begin{tabular}{|l|cccccc|cc|}
\hline
\multicolumn{1}{|c|}{}  & \multicolumn{1}{c}{T$_0$} & \multicolumn{1}{c}{T$_{90}$} & \multicolumn{1}{c}{z}  & \multicolumn{1}{c}{Start time} &  \multicolumn{1}{c}{Zenith} & \multicolumn{1}{c|}{Delay} & \multicolumn{1}{c}{Trigger} & \multicolumn{1}{c|}{VHE} \\
  & [UTC] & [s] &  & [UTC] & [deg.] & [s] & &  \\
\hline
GRB\,201216C & 23:07:31 & 48.0 & 1.1 & 20:57:03 & 40 & 79200 &  {\it Swift} & Y$^{\alpha}$\\
GRB\,210217A & 23:25:42 & 4.2 & - & 23:40:22 & 44 & 880 & {\it Swift} & N  \\
GRB\,210511B & 11:26:39 & 6  & - & 03:37:54 & 45 & 58200 & {\it Fermi}-GBM & N \\
\hline
IC\,210210A & 11:53:55 & - & - & 05:41:54 & 25 & 64134 & IceCube & N \\
\hline
\end{tabular}
\caption[Transient alerts observed by LST-1]%
{Transient follow-up observed by LST-1. Columns represent respectively: the transient name, the satellite trigger time, the duration of the event at X-rays (T$_{90}$), the GRB redshift, the start time of LST-1 observation, the zenith angle at the beginning of the follow-up and the overall delay between the beginning of data taking and the burst onset. Last two columns represent the instrument that provide the trigger and a yes/no flag for a detected VHE counterpart. $^{(\alpha)}$ from MAGIC} \label{tab:transients}
\end{center}
\end{table}

The preliminary data analysis for the events in Tab.~\ref{tab:transients} was performed using the LST-1 data analysis package \texttt{lst-chain}. For the gamma/hadron separation, a multivariate method based on a random forest (RF) algorithm was applied. This algorithm employs some Cherenkov image parameters ~\cite{hil85} to compute a gamma/hadron discriminator called \emph{gammaness} by comparison with Monte Carlo gamma-ray simulations. The detection of the possible gamma-ray signal is achieved through the so-called $\theta^2$ plot, i.e. the comparison between the distributions of the squared angular distance between the reconstructed position of the source and its nominal position in the signal and background regions for energies above the threshold. The significance of the signal is evaluated using single cuts in \emph{gammaness} and $\theta^2$ and according to Eq.~17 of~\cite{lima83}. Preliminary results did not reveal any significant VHE emission above the energy threshold for any of the observed events.

\begin{figure}
\centering
\includegraphics[width=0.48\columnwidth]{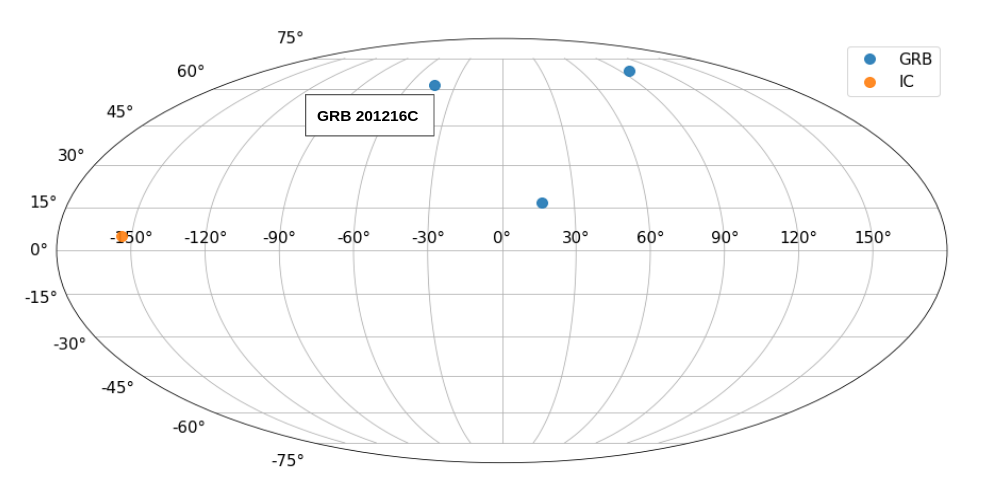}
\includegraphics[width=0.45\columnwidth]{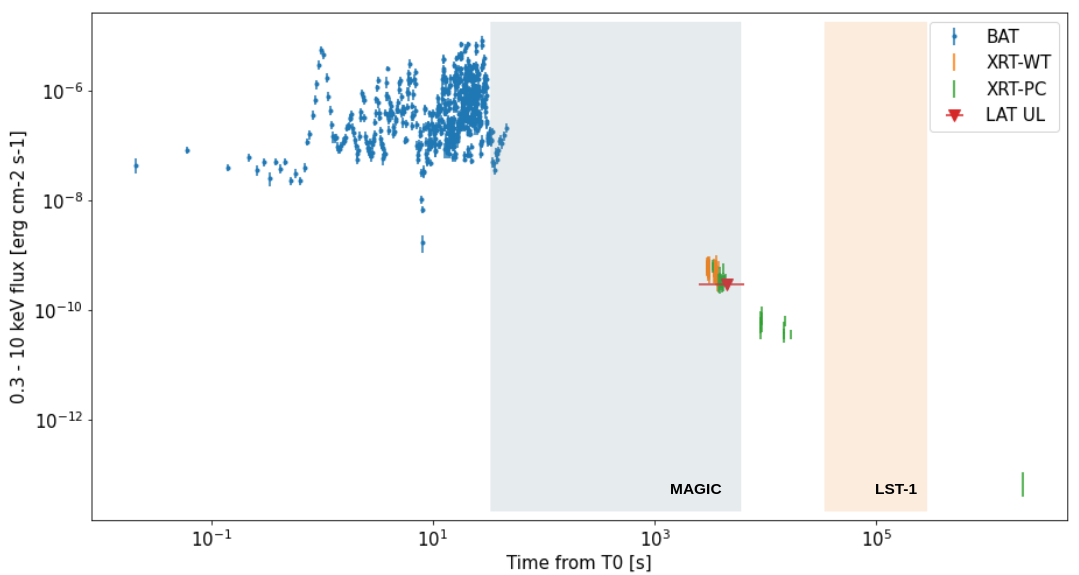}
\caption{{\it (Left panel):} Aitoff projection for the transient alerts observed so far by LST-1. {\it (Right panel):} 0.3-10 keV light curve for GRB\,201216C as measured by the combined BAT and XRT instruments of the {\it Swift} satellite. {\it Fermi}-LAT upper limits is also reported~\cite{lat201216c}.}
\label{fig:201216c}
\end{figure}

\subsection{GRB\,201216C}
GRB\,201216C was triggered and located at 23:07:31 UT by {\it Swift}-BAT. The BAT light curve shows a multi-peaked structure between T$_0$-16 s and T$_0$+64 s for a duration of T$_{90} \sim$ 48 s. The 0.3-10 keV light curve with MAGIC and LST-1 observation window is reported in Fig.~\ref{fig:201216c}. GRB\,201216C is a relatively (for GRBs) low redshift event ($z=1.1$). With an equivalent isotropic energy release of $\sim 6.2 \times 10^{53}$ erg and a rest-frame peak energy of $700 \pm 61$ keV~\cite{GCN29084}, the event is in line with the Amati relation and close to the parameter space location of other VHE detected GRB like GRB\,180720B and GRB\,190114C~\ref{fig:amati}. Indeed, GRB\,201216C represents a particularly interesting case as the event was detected at VHE by the MAGIC telescopes~\cite{GCN29075} that was able to point at GRB coordinates in less then 1 minute with respect to T$_0$. Triggered by the VHE detection, the event was followed up by LST-1 the day after (at the time of the alert, the transient handler was not yet fully operational) starting the observation at a moderate zenith angle ($40^{\circ}$). Unfortunately, no significant emission was detected.   

\section{Conclusions}
The LST prototype, the largest IACT foreseen for the CTA, is finalizing its commissioning phase and gradually increasing the hours dedicated to scientific observations. In this phase, a dedicated transient handler has been developed to allow the follow-up of transient alerts. During the first months of 2021, some initial follow-up observations have been performed. Being in commissioning phase, these results have to be considered preliminary and deriving from a yet-not-automatic and optimized observation procedure. Furthermore, dedicated analysis optimization is currently under development. Preliminary analysis did not reveal VHE signal associated to any of the observed alerts. However, the rising number of GRBs detected at VHE and the importance of coordinated efforts for follow-up observations of multi-messenger alerts will soon put LST-1 in a key position for VHE observations of those peculiar events; in particular, considering the possible joined observations with the MAGIC telescopes.

%% Full authors list (ONLY FOR COLLABORATIONS)
%\clearpage
\section*{Full Authors List: \Coll\ Collaboration}
\scriptsize
\noindent
H. Abe$^{1}$,
A. Aguasca$^{2}$,
I. Agudo$^{3}$,
L. A. Antonelli$^{4}$,
C. Aramo$^{5}$,
T.  Armstrong$^{6}$,
M.  Artero$^{7}$,
K. Asano$^{1}$,
H. Ashkar$^{8}$,
P. Aubert$^{9}$
% ...
\\

\noindent
$^{1}$Institute for Cosmic Ray Research, University of Tokyo.
$^{2}$Departament de Física Quàntica i Astrofísica, Institut de Ciències del Cosmos, Universitat de Barcelona, IEEC-UB.
$^{3}$Instituto de Astrofísica de Andalucía-CSIC.
$^{4}$INAF - Osservatorio Astronomico di Roma.
$^{5}$INFN Sezione di Napoli.
$^{6}$Aix Marseille Univ, CNRS/IN2P3, CPPM.
$^{7}$Institut de Fisica d'Altes Energies (IFAE), The Barcelona Institute of Science and Technology.
$^{8}$IRFU, CEA, Université Paris-Saclay.
$^{9}$LAPP, Univ. Grenoble Alpes, Univ. Savoie Mont Blanc, CNRS-IN2P3, Annecy.
% ...

%\noindent \textbf{Note comment afterwards:} Collaborations have the possibility to provide an authors list in xml format which will be used while generating the DOI entries making the full authors list searchable in databases like Inspire HEP. For instructions please go to icrc2021.desy.de/proceedings or contact us under icrc2021proc@desy.de.\\
%
%\scriptsize
%\noindent
%first.author$^1$, 
%second.author$^2$, 
%third.author$^3$ % .... more names
%and 
%last.author$^{n}$ \\
%
%\noindent
%$^1$first.affiliation.
%$^2$second.affiliation. % .... more affiliation
%$^{m}$last.affiliation.

\end{document}